# Calculation of permittivity tensors for invisibility devices by effective medium approach in general relativity

Doyeol Ahn*

*Center for Quantum Information Processing, Department of Electrical and Computer Engineering, University of Seoul, Seoul 130-743, Republic of Korea*

**Abstract:**

Permittivity tensors of arbitrary shaped invisibility devices are obtained using effective medium approach in general relativity. As special cases, analytical expressions for the permittivity tensors of invisibility cloaks for the elliptic cylinder, prolate spheroid, and the confocal paraboloid geometry are derived. In the case of elliptic cylinder, we found that the point of infinite light speed in the electromagnetic space becomes two points in the physical space in the zz component of the permittivity tensor. This result is different from the case of perfect cylinder in which there is a line of cloak at which the speed of light becomes infinite. In the cases of prolate spheroid and confocal paraboloid, the point of infinite light speed in the electromagnetic space becomes line in the physical space for the first two tensor components and the third component of the permittivity tensor becomes singular at the line of cloak.

*To whom correspondence should be addressed. E-mail: dahn@uos.ac.kr

# 1. Introduction

Control of electromagnetic fields using meta-materials for novel applications such as invisibility devices, negative refraction and artificial black holes has been an exciting development [1-6]. Based on coordinate transformations and conformal mappings of Maxwell's equations, Pendry et al. [3] and Leonhardt [4], independently, proposed an invisibility cloak which hides objects inside of given geometrical shape. When electromagnetic waves pass through the invisibility device, it will deflect the waves, guide them around the object and return the waves to the original propagation path without disturbing the external fields. This phenomenon has been verified experimentally for the cylindrical form cloak [7].

The most important issue in the analysis and design of the invisibility devices is the calculation of permittivity and permeability tensors for meta-materials to fill in the cloak shell. Assume that we might start with a uniform field lines then we would like to let the invisibility device distort the field lines be moved to avoid certain region. This distortion can be regarded as a coordinate transformation between the original Cartesian mesh and the distorted mesh [3]. Conventional approach is to keep the form of Maxwell's equations invariant in any coordinate but the permittivity tensor $\varepsilon_{ij}$ and the permeability tensor $\mu_{ij}$ are scaled by the factors obtained by the general coordinate transformation [3] or by the optical conformal mapping technique [4]. There have been both numerical and analytical results for elliptical shapes [8-10] and more complicated geometries [11-15] using above mentioned techniques.

An alternative approach to calculate the permittivity and permeability tensors for invisibility devices using electrodynamics in the frame of general relativity has been proposed [6,16]. This approach is based on the discovery due to Tamm [17] who noted that the propagation of classical electromagnetic waves in the presence of curved spacetime could be formally case into the flat spacetime with the dielectric and magnetic medium. The original idea of Tamm has been extended to study the deflection and rotation of light in a non-uniform gravitational field [18], the negative phase velocity in vacuum[19,20] and meta-materials [21], and negative refraction in classical vacuum [22,23]. The principal idea is based on the fact that propagation of electromagnetic waves in curved space-time can formally be described as light traveling in an inhomogeneous effective bi-anisotropic medium whose constitutive parameters are determined by the space-time metric [16-23]. Then one can formulate the inverse problem of transforming a medium in flat spacetime into some curved spacetime and find specific conditions for the invisibility cloaking. The advantage of this approach is that both permittivity and permeability tensors can be obtained both intuitively and directly for both two- and three-dimensional invisibility devices. However, detailed calculations except for the simple cylindrical and spherical shapes are not given yet. In this work, the author presents the transformation rules for cloaking arbitrary shaped objects based on the effective medium approach for Maxwell's equations in general relativity. As special cases, analytical expressions for the permittivity tensors of invisibility cloaks for the elliptic cylinder, prolate spheroid, and the confocal paraboloid geometry are derived.

## 2. Effective medium approach in general relativity

In this section, we first review the derivation of permittivity and permeability tensors using effective medium approach [6, 16-23] in general relativity and extend the formulation to obtain the transformation rules for arbitrary shaped invisibility devices. In order to determine the effects of gravitation or curved space-time on a general physical system, we replace all Lorentz tensors, which describe the given special-relativistic equations in Minkowski spacetime, with objects that behave like tensors under general coordinate transformations [24,25] Also, we replace all derivatives with covariant derivatives and replace the Minkowski metric tensor $\eta_{ab}$ with the metric tensor $g_{\mu\nu}$, where $\eta_{ab}$ is the Minkowski metric tensor with the following components:

$$\eta_{00} = -1, \ \eta_{11} = \eta_{22} = \eta_{33} = 1. \tag{1}$$

Then the equations are then generally covariant. Covariant Maxwell's equations in general relativity are given by [16-25]

$$F^{\mu\nu}{}_{;\mu} = \frac{\varepsilon_0}{\sqrt{-g}} \frac{\partial}{\partial x^\mu}\left(\sqrt{-g} F^{\mu\nu}\right) = -J^\nu \qquad (2)$$

and

$$F_{\mu\nu;\lambda} + F_{\lambda\mu;\nu} + F_{\nu\lambda;\mu} = 0, \qquad (3)$$

where the electromagnetic field tensor $F_{\mu\nu}$ is defined as

$$F_{\mu\nu} = \begin{pmatrix} 0 & -E_x & -E_y & -E_z \\ E_x & 0 & B_z & -B_y \\ E_y & -B_z & 0 & B_x \\ E_z & B_y & -B_x & 0 \end{pmatrix}, \qquad (4)$$

$g$ is the determinant of the metric tensor $g_{\mu\nu}$, $\varepsilon_0$ is the permittivity of free space and we set $c = \hbar = G = 1$.

We also define a new contra-variant tensor $H^{\mu\nu}$ as [22]

$$H^{\mu\nu} = \varepsilon_o \frac{\sqrt{-g}}{2}\left(g^{\mu\lambda}g^{\nu\rho} - g^{\mu\rho}g^{\nu\lambda}\right)F_{\lambda\rho} \qquad (5)$$

with

$$H^{\mu\nu} = \begin{pmatrix} 0 & D_x & D_y & D_z \\ -D_x & 0 & H_z & -H_y \\ -D_y & -H_z & 0 & H_x \\ -D_z & H_y & -H_x & 0 \end{pmatrix}. \qquad (6)$$

From equations (2) to (6), we get [16]:

$$D_i = (-g)^{1/2}\varepsilon_0\left(g^{0j}g^{i0} - g^{00}g^{ij}\right)E_j + (-g)^{1/2}[jkl]g^{0k}g^{il}\mu_0^{-1}B_j , \qquad (7)$$

and

$$H_i = \frac{1}{\sqrt{-g}}[jkl]g_{ok}g_{il}\varepsilon_0 E_j + \frac{1}{\sqrt{-g}}\left(g_{io}g_{jo} - g_{00}g_{ij}\right)\mu_0^{-1}B_j \qquad (8)$$

where $[ijk]$ is the completely anti-symmetric permutation symbol with $[xyz] = 1$.

Equations (7) and (8) can be rewritten in more convenient forms [22]:

$$D_i = \varepsilon^{ij}\varepsilon_0 E_j + \alpha^{ij}\mu_0^{-1}B_j . \qquad (9)$$

and

$$B_i = \beta_{ij}\varepsilon_0 E_j + (\mu^{-1})_{ij}\mu_0^{-1}B_j , , \qquad (10)$$

with the symmetric tensors are given by (Appendix)

$$\begin{aligned}\varepsilon^{ij} &= (-g)^{1/2}\left(g^{0j}g^{i0} - g^{00}g^{ij}\right)\\ \alpha^{ij} &= (-g)^{1/2}[jkl]g^{0k}g^{il},\end{aligned} \qquad (11)$$

and

$$\begin{aligned}\beta_{IJ} &= \frac{1}{\sqrt{-g}}[jkl]g_{0k}g_{il},\\ (\mu^{-1})_{ij} &= \frac{1}{\sqrt{-g}}\left(g_{i0}g_{j0} - g_{00}g_{ij}\right) .\end{aligned} \qquad (12)$$

Detailed derivation of equations (9) to (12) from equations (7) and (8) is given in the appendix. From this, we can see that empty curved-space-time appears inhomogeneous effective bi-anisotropic medium whose electric permittivity and magnetic permeability tensors are determined by the space-time metric.

Conversely, one might be able to describe the dielectric medium by some kind of coordinate transformation from the empty but curved space-time or curved coordinate. The inverse metric tensor is transformed as

$$g^{\alpha\beta} = \frac{\partial x^\alpha}{\partial x'^\mu} \frac{\partial x^\beta}{\partial x'^\nu} g'^{\mu\nu}, \tag{13}$$

and the covariant metric tensor is transformed as

$$g_{\alpha\beta} = \frac{\partial x'^\mu}{\partial x^\alpha} \frac{\partial x'^\nu}{\partial x^\beta} g'_{\mu\nu}. \tag{14}$$

We assume that the physical medium can be described by the curved coordinates $x^i$ with spatial metric $\gamma_{ij}$ and the determinant $\gamma$. The metric $\gamma_{ij}$ should be different from the spatial part of the effective geometry $g_{\alpha\beta}$ generated by the medium because, in this picture, the metric $g_{\alpha\beta}$ does not describe the actual space-time but effective geometry corresponding to the original bi-anisotropic medium while $\gamma_{ij}$ describes the actual spatial coordinate.

Since we are mainly interested in the implementation of invisibility devices in the three dimensional space, we focus on the spatial transformation media which perform spatial coordinate transformation of electromagnetic field. Let's assume that the right handed Cartesian coordinates x, y, and z are related to the curvilinear coordinates $x^1, x^2$ and $x^3$ by

$$\begin{aligned} x &= x(x^1, x^2, x^3), \\ y &= y(x^1, x^2, x^3), \\ z &= z(x^1, x^2, x^3), \end{aligned} \tag{15}$$

and by the metric

$$ds^2 = dx^2 + dy^2 + dz^2 = \sum_{i,j} \gamma_{ij} dx^i dx^j .$$ (16)

For simplicity, we would like to hide any object in the region specified by the curvilinear coordinates $x^1, x^2, x^3$: $0 < x^1 < U_1, 0 < x^2 < V_1, 0 < x^3 < W_1$ and the invisibility device is consisting of meta-material shell in $U_1 < x^1 < U_2, V_1 < x^2 < V_2, W_1 < x^3 < W_2$. We use primed coordinate for the empty curved space-time and define the physical medium by

$$x^1 = U_1 + \frac{U_2 - U_1}{U_2} x'^1,$$
$$x^2 = V_1 + \frac{V_2 - V_1}{V_2} x'^2,$$ (17)
$$x^3 = W_1 + \frac{W_2 - W_1}{W_2} x'^3,$$

and define the effective geometry corresponding to the original bi-anisotropic medium by

$$g^{ij} = \frac{\partial x^i}{\partial x'^k} \frac{\partial x^j}{\partial x'^l} \gamma'^{kl}$$ (18)

and

$$\sqrt{-g} = \sqrt{-g'} \frac{\partial(x'^1, x'^2, x'^3)}{\partial(x^1, x^2, x^3)} = \sqrt{\gamma'} \frac{\partial(x'^1, x'^2, x'^3)}{\partial(x^1, x^2, x^3)} .$$ (19)

Here, we have $\gamma = \det(\gamma_{ij})$ and $\gamma^{kk} = 1/\gamma_{kk}$.

Equations (15) to (19) give the transformation formula for the calculation of permittivity and permeability tensors. By taking into account of the spatially covariant forms of divergences in the Maxwell's equations, the constitutive parameters are then given by [16,19,22]:

$$\varepsilon^{ij} = \pm \frac{(-g)^{1/2}}{\sqrt{\gamma}} \left( g^{0j} g^{i0} - g^{00} g^{ij} \right)$$

$$\alpha^{ij} = \frac{(-g)^{1/2}}{\sqrt{\gamma}} [jkl] g^{0k} g^{il},$$

(20)

and

$$\beta_{IJ} = \frac{\sqrt{\gamma}}{\sqrt{-g}} [jkl] g_{0k} g_{il},$$

$$\left( \mu^{-1} \right)_{ij} = \pm \frac{\sqrt{\gamma}}{\sqrt{-g}} \left( g_{i0} g_{j0} - g_{00} g_{ij} \right).$$

(21)

Using $g_{\mu\lambda} g^{\lambda\nu} = \delta^{\nu}_{\mu}$, we obtain simplified expressions for the permittivity and permeability tensors as

$$\varepsilon^{ij} = \mu^{ij} = \mp \frac{\sqrt{-g}}{g_{00} \sqrt{\gamma}} g^{ij},$$

(22)

where the minus sign indicates the medium with negative refraction.

Suppose the transformed space from the original empty curved space-time does not cover the entire physical space for the medium and the medium excludes electromagnetic fields in certain regions but smoothly fits them to fields outside the device. Therefore electromagnetic radiation is guided around the excluded regions. As a result, the medium cloaks these regions such that any object placed inside is hidden without revealing anything to the outside the world. Invisibility device should employ anisotropic media because the inverse scattering problem for waves in isotropic media has unique solutions. The implementation of the invisibility device or the radiation shield employs the coordinate transformation with the holes.

An alternative approach would be to use 'natural invariance' as first shown by Post [26, 27] in which all connection terms vanish and covariant derivatives can be replaced by the ordinary derivatives. For example, if we introduce a new tensor $T^{\lambda\nu} = \frac{1}{2}\chi^{\lambda\nu\sigma\kappa}F_{\sigma\kappa}$ where $\chi^{\lambda\nu\sigma\kappa}$ is called constitutive tensor and a potential vector $A_\nu$ according to $F_{\lambda\nu} = \partial_\lambda A_\nu - \partial_\nu A_\lambda$, equations (2) and (3) can be replaced by following equations with ordinary derivatives; $\partial_\nu T^{\lambda\nu} = 0$ and $\partial_\nu \chi^{\lambda\nu\sigma\kappa}\partial_\sigma A_\kappa = 0$.

Once the spatial geometry or curvature is specified, the field can be calculated by the generalized Helmholtz equation [28]

$$(-g)^{1/2}\frac{\partial}{\partial x^\mu}\left[g^{\mu\nu}(-g)^{1/2}\frac{\partial}{\partial x^\nu}\right]\psi = 0. \tag{23}$$

### 3. Analytical calculation of permittivity tensors

In this section, we focus our attention to the calculation of permittivity tensors and we consider following non-trivial geometrical shapes for the invisibility devices. For simplicity, we consider the cases with both $V_1 = 0$ and $W_1 = 0$ in equation (17).

3.1. Elliptical cylinder

We consider the elliptic cylindrical space-time where the coordinates are given by

$$\begin{aligned} x &= \cosh u \cos v \\ y &= \sinh u \sin v \\ z &= z \end{aligned} \tag{24}$$

and the spatial metric is given by

$$\gamma_{ij} = \begin{pmatrix} \sinh^2 u + \sin^2 v & 0 & 0 \\ 0 & \sinh^2 u + \sin^2 v & 0 \\ 0 & 0 & 1 \end{pmatrix} \tag{25}$$

and $\gamma = \left(\sinh^2 u + \sin^2 v\right)^2$.

For an invisibility device, z-axis is stretched out to a full elliptic. Assume that we would like to hide any object in the region $0 < u < U_1$ (Fig. 1) and the invisibility device is consisting of meta-material shell in $U_1 < u < U_2$.

Let's use primed coordinate for the empty curved space-time and define the physical medium by

$$u = U_1 + u' \frac{U_2 - U_1}{U_2}$$
$$v = v' \tag{26}$$
$$z = z'$$

We also have

$$\frac{\partial u}{\partial u'} = \frac{U_2 - U_1}{U_2}, \tag{27}$$

$$g^{ij} = \frac{\partial x^i}{\partial x'^k} \frac{\partial x^j}{\partial x'^l} \gamma'^{kl}$$

$$= \begin{pmatrix} \left(\frac{U_2 - U_1}{U_2}\right)^2 \frac{1}{\sinh^2 u' + \sin^2 v'} & 0 & 0 \\ 0 & \frac{1}{\sinh^2 u' + \sin^2 v'} & 0 \\ 0 & 0 & 1 \end{pmatrix}, \tag{28}$$

and

$$\sqrt{-g} = \sqrt{-g'} \frac{\partial(x'^1, x'^2, x'^3)}{\partial(x^1, x^2, x^3)} = (\sinh^2 u' + \sin^2 v') \frac{U_2}{U_2 - U_1}. \tag{29}$$

By the way

$$\sqrt{\gamma} = \sinh^2 u + \sin^2 v. \tag{30}$$

From, equations (20), (28) to (30), we get

$$\varepsilon^{ij} = \frac{1}{\sinh^2 u + \sin^2 v} \begin{pmatrix} \frac{U_2 - U_1}{U_2} & 0 & 0 \\ 0 & \frac{U_2}{U_2 - U_1} & 0 \\ 0 & 0 & \frac{U_2}{U_2 - U_1}(\sinh^2 u' + \sin^2 v') \end{pmatrix}. \quad (31)$$

Or in terms of mixed tensor notation, we have

$$\varepsilon^i_j = \varepsilon^{ik} \gamma_{kj} \quad (32)$$

$$= \begin{pmatrix} \frac{U_2 - U_1}{U_2} & 0 & 0 \\ 0 & \frac{U_2}{U_2 - U_1} & 0 \\ 0 & 0 & \frac{U_2}{U_2 - U_1} \frac{(\sinh^2 u' + \sin^2 v')}{(\sinh^2 u + \sin^2 v)} \end{pmatrix}.$$

The mixed tensor $\varepsilon^i_j$ is sometimes convenient to use because the eigenvalues of $\varepsilon^i_j$ are invariant under coordinate transformations and consequently one can read off the dielectric properties from the eigenvalues of $\varepsilon^i_j$ [5].

Fig. 2 shows the permittivity tensor distribution in the physical space for the z = 0 plane. We assume that $U_1 = 0.1$ and $U_2 = 0.2$. In equations (31) and (32), close to the lining of the cloak at $u \to U_1$ where $u' \to 0$, the z-component of the permittivity tensor becomes zero only when $v = 0$ and $v = \pi$ as a result the speed of light diverges [4, 16]. In electromagnetic space, electromagnetic waves pass a point in infinity and in physical space the single point becomes two points. This result is different from the case of perfect cylinder in which there is a line of cloak at which the speed of light becomes infinite. This would allow more flexible design of meta-materials. It is interesting to note that the first two components of the mixed permittivity tensors are constant and the zz-component is the same as the contravariant tensor.

3.2. Prolate spheroid

In this example, we consider the prolate spheroidal space-time where the coordinates are given by

$$x = \sinh u \sin v \cos \varphi$$
$$y = \sinh u \sin v \sin \varphi \qquad (33)$$
$$z = \cosh u \cos v$$

and the spatial metric is given by

$$\gamma_{ij} = \begin{pmatrix} \sinh^2 u + \sin^2 v & 0 & 0 \\ 0 & \sinh^2 u + \sin^2 v & 0 \\ 0 & 0 & \sinh^2 u \sin^2 v \end{pmatrix} \qquad (34)$$

and $\gamma = \sinh^2 u \sin^2 v (\sinh^2 u + \sin^2 v)^2$.

Assume that we would like to hide any object in the region $0 < u < U_1$ (Fig. 3) and the invisibility device is consisting of meta-material shell in $U_1 < u < U_2$. Let's us primed coordinate for the empty curved space-time and define the physical medium by

$$u = U_1 + u' \frac{U_2 - U_1}{U_2}$$
$$v = v' \qquad (35)$$
$$\varphi = \varphi'$$

We also have

$$\frac{\partial u}{\partial u'} = \frac{U_2 - U_1}{U_2}, \qquad (36)$$

$$g^{ij}$$

$$= \frac{\partial x^i}{\partial x'^k}\frac{\partial x^j}{\partial x'^l}\gamma'^{kl} \quad , \tag{37}$$

$$= \begin{pmatrix} \left(\dfrac{U_2-U_1}{U_2}\right)^2 \dfrac{1}{\sinh^2 u'+\sin^2 v'} & 0 & 0 \\ 0 & \dfrac{1}{\sinh^2 u'+\sin^2 v'} & 0 \\ 0 & 0 & \dfrac{1}{\sinh^2 u'\sin^2 v'} \end{pmatrix}$$

and

$$\sqrt{-g} = \sqrt{-g'}\,\frac{\partial(x'^1,x'^2,x'^3)}{\partial(x^1,x^2,x^3)} = (\sinh^2 u'+\sin^2 v')\sinh u'\sin v'\,\frac{U_2}{U_2-U_1}. \tag{38}$$

By the way

$$\sqrt{\gamma} = (\sinh^2 u + \sin^2 v)\sinh u \sin v. \tag{39}$$

From, equations (20), (36) to (39), we get

$$\varepsilon^{ij} = \begin{pmatrix} \dfrac{U_2-U_1}{U_2}\sinh u'\sin v' & 0 & 0 \\ 0 & \dfrac{U_2}{U_2-U_1}\sinh u'\sin v' & 0 \\ 0 & 0 & \dfrac{U_2}{U_2-U_1}\dfrac{(\sinh^2 u'+\sin^2 v')}{\sinh u'\sin v'} \end{pmatrix}, \tag{40}$$

$$\times \frac{1}{(\sinh^2 u+\sin^2 v)\sinh u \sin v}$$

and

$$\varepsilon^i_j$$
$$= \varepsilon^{ik}\gamma_{kj} \tag{41}$$
$$= \begin{pmatrix} \dfrac{U_2-U_1}{U_2}\dfrac{\sinh u'\sin v'}{\sinh u \sin v} & 0 & 0 \\ 0 & \dfrac{U_2}{U_2-U_1}\dfrac{\sinh u'\sin v'}{\sinh u \sin v} & 0 \\ 0 & 0 & \dfrac{U_2}{U_2-U_1}\dfrac{\sinh u\sin v(\sinh^2 u'+\sin^2 v')}{\sinh u'\sin v'(\sinh^2 u+\sin^2 v)} \end{pmatrix}.$$

Fig. 4 shows the permittivity tensor distribution in the physical space for the z = 0 plane. We assume that $U_1 = 0.1$ and $U_2 = 0.2$. In this case, close to the lining of the cloak at $u \to U_1$ where $u' \to 0$ $\varepsilon^{uu}$ and $\varepsilon^{vv}$ components of the permittivity tensor have line of infinite light velocity in physical space but $\varepsilon^{\varphi\varphi}$ component becomes singular. As a result, it is expected that the phase velocity of the $\varphi$ component is expected to be zero. Fig. 5 shows the mixed tensor $\varepsilon^i_j$ distribution in the physical space for the z=0 plane. We note that $\varepsilon^u_u$ and $\varepsilon^v_v$ show similar distribution both in shape and scale with $\varepsilon^{uu}$ and $\varepsilon^{vv}$, respectively, but $\varepsilon^\varphi_\varphi$ shows reduced scale when compared with that of $\varepsilon^{\varphi\varphi}$.

### 3.3. Confocal parabolid

As an example, we consider the confocal parabolid space-time where the coordinates are given by

$$\begin{aligned} x &= \xi\eta\cos\varphi \\ y &= \xi\eta\sin\varphi \\ z &= \frac{1}{2}(\eta^2 - \xi^2) \end{aligned} \tag{42}$$

and the spatial metric is given by

$$\gamma_{ij} = \begin{pmatrix} \xi^2+\eta^2 & 0 & 0 \\ 0 & \xi^2+\eta^2 & 0 \\ 0 & 0 & (\xi\eta)^2 \end{pmatrix} \tag{43}$$

and $\gamma = \xi^2\eta^2(\xi^2+\eta^2)^2$.

Assume that we would like to hide any object in the region $0 < u < U_1$ (Fig. 6) and the invisibility device is consisting of meta-material shell in $U_1 < u < U_2$.

Let's us primed coordinate for the empty curved space-time and define the physical medium by

$$\xi = U_1 + \xi' \frac{U_2 - U_1}{U_2}$$
$$\eta = \eta' \tag{44}$$
$$\varphi = \varphi'$$

We also have

$$\frac{\partial \xi}{\partial \xi'} = \frac{U_2 - U_1}{U_2}, \tag{45}$$

$$g^{ij} = \frac{\partial x^i}{\partial x'^k} \frac{\partial x^j}{\partial x'^l} \gamma'^{kl}$$

$$= \begin{pmatrix} \left(\frac{U_2 - U_1}{U_2}\right)^2 \frac{1}{\xi'^2 + \eta'^2} & 0 & 0 \\ 0 & \frac{1}{\xi'^2 + \eta'^2} & 0 \\ 0 & 0 & \frac{1}{\xi'^2 \eta'^2} \end{pmatrix}, \tag{46}$$

and

$$\sqrt{-g} = \sqrt{-g'} \frac{\partial(x'^1, x'^2, x'^3)}{\partial(x^1, x^2, x^3)} = (\xi'^2 + \eta'^2) \xi' \eta' \frac{U_2}{U_2 - U_1}. \tag{47}$$

By the way

$$\sqrt{\gamma} = (\eta^2 + \xi^2) \xi \eta. \tag{48}$$

From, equations (15), (38) to (42), we get

$$\varepsilon^{ij} = \begin{pmatrix} \dfrac{U_2 - U_1}{U_2}\xi'\eta' & 0 & 0 \\ 0 & \dfrac{U_2}{U_2 - U_1}\xi'\eta' & 0 \\ 0 & 0 & \dfrac{U_2}{U_2 - U_1}\dfrac{(\xi'^2 + \eta'^2)}{\xi'\eta'} \end{pmatrix},$$
$$\times \dfrac{1}{(\xi^2 + \eta^2)\xi\eta}$$
(49)

and

$$\varepsilon_j^i = \varepsilon^{ik}\gamma_{kj}$$
$$= \begin{pmatrix} \dfrac{U_2 - U_1}{U_2}\dfrac{\xi'\eta'}{\xi\eta} & 0 & 0 \\ 0 & \dfrac{U_2}{U_2 - U_1}\dfrac{\xi'\eta'}{\xi\eta} & 0 \\ 0 & 0 & \dfrac{U_2}{U_2 - U_1}\dfrac{(\xi'^2 + \eta'^2)\xi\eta}{(\xi^2 + \eta^2)\xi'\eta'} \end{pmatrix}.$$
(50)

Fig. 7 shows the permittivity tensor distribution in the physical space for the z = 0 plane. We assume that $U_1 = 0.1$ and $U_2 = 0.2$. In this case, close to the lining of the cloak at $\xi \to U_1$ where $\xi' \to 0$, $\varepsilon^{\xi\xi}$ and $\varepsilon^{\eta\eta}$ components of the permittivity tensor have line of infinite light velocity in physical space but $\varepsilon^{\varphi\varphi}$ component becomes singular. As a result, it is expected that the phase velocity of the $\varphi$ component is expected to be zero. Fig. 8 shows the mixed tensor $\varepsilon_j^i$ distribution in the physical space for the z=0 plane. We note that $\varepsilon_\xi^\xi$ and $\varepsilon_\eta^\eta$ show similar distribution both in shape and scale with $\varepsilon^{\xi\xi}$ and $\varepsilon^{\eta\eta}$, respectively, but $\varepsilon_\varphi^\varphi$ shows reduced scale when compared with that of $\varepsilon^{\varphi\varphi}$.

## 4. Summary

In this work, we derived permittivity and permeability tensors for arbitrary shaped invisibility devices and analytical expressions for the permittivity tensors of invisibility cloaks for the cases of elliptic cylinder, prolate spheroid, and the confocal paraboloid as special cases using relations between the electromagnetic tensor and its dual tensor and their application to the transformation between the electromagnetic spacetime and the physical spacetime. In the case of elliptic cylinder, we found that the point of infinite light speed in the electromagnetic space becomes two points in the physical space. This result is different from the case of perfect cylinder in which there is a line of cloak at which the speed of light becomes infinite. In the cases of prolate spheroid and confocal paraboloid, the point of infinite light speed in the electromagnetic space becomes line in the physical space for the first two tensor components and the third component of the permittivity tensor becomes singular at the line of cloak.

**Acknowledgements**

This work was supported by the University of Seoul through the Advanced Research Facility Grant of 2010 by the Seoul Metropolitan government. The author thanks Dr. J. H Oh for his help with the plots.

**Appendix: Derivation of equations (9)-(12)**

We start with the electromagnetic field tensor $F_{\mu\nu}$ which is defined as

$$F_{\mu\nu} = \begin{pmatrix} 0 & -E_x & -E_y & -E_z \\ E_x & 0 & B_z & -B_y \\ E_y & -B_z & 0 & B_x \\ E_z & B_y & -B_x & 0 \end{pmatrix}, \tag{A1}$$

and we define a contra-variant tensor $H^{\alpha\beta}$ by

$$H^{\alpha\beta} = \frac{1}{2} \chi^{\alpha\beta\mu\nu} F_{\mu\nu}, \tag{A2}$$

with $\quad \chi^{\alpha\beta\mu\nu} = (-g)^{1/2} \varepsilon_0 \left( g^{\alpha\mu} g^{\beta\nu} - g^{\alpha\nu} g^{\beta\mu} \right),$ (A3)

where $\varepsilon_0$ is the permittivity of the free space. In this work, we set the speed of light $c = 1$, so the vacuum permeability $\mu_0$ is related to $\varepsilon_0$ by the relation $\varepsilon_0 \mu_0 = 1/c^2 = 1$.

From (A2) and (A3), we obtain

$$H^{\alpha\beta} = \frac{1}{2}(-g)^{1/2}\varepsilon_0\left(g^{\alpha\mu}g^{\beta\nu} - g^{\alpha\nu}g^{\beta\mu}\right)F_{\mu\nu}$$
$$= (-g)^{1/2}\varepsilon_0 g^{\alpha\mu}g^{\beta\nu},$$
(A4)

where we have used the anti symmetric property $F_{\mu\nu} = -F_{\nu\mu}$ of the electromagnetic field tensor. If we define $D_i = H^{0i}$, then from (A2) to (A4)

$$\begin{aligned}D_i &= (-g)^{1/2}\varepsilon_0 g^{0\mu}g^{i\nu}F_{\mu\nu} \\ &= (-g)^{1/2}\varepsilon_0 g^{00}g^{i\nu}F_{0\nu} + (-g)^{1/2}\varepsilon_0 g^{0k}g^{i\nu}F_{k\nu} \\ &= (-g)^{1/2}\varepsilon_0\left(g^{0j}g^{i0} - g^{00}g^{ij}\right)E_j + (-g)^{1/2}[jkl]g^{0k}g^{il}\mu_0^{-1}B_j \\ &= \varepsilon^{ij}\varepsilon_0 E_j + \alpha^{ij}\mu_0^{-1}B_j.\end{aligned}$$
(A5)

Here $[ijk]$ is the completely anti-symmetric permutation symbol with $[xyz] = 1$ and

$$\varepsilon^{ij} = (-g)^{1/2}\left(g^{0j}g^{i0} - g^{00}g^{ij}\right)$$
$$\alpha^{ij} = (-g)^{1/2}[jkl]g^{0k}g^{il}.$$
(A6)

It is convenient to define the dual tensor $^*H_{\mu\nu}$ to derive the expression for the permeability tensor by

$$^*H_{\mu\nu} = \frac{1}{2}\varepsilon_{\mu\nu\lambda\rho}H^{\lambda\rho}$$
(A7)

where four dimensional Levi Civita tensor is defined by [29,30]

$$\varepsilon_{\mu\nu\lambda\rho} = \sqrt{-g}[\mu\nu\lambda\rho]$$
(A8)

with $[0123] = 1$. Then it is straight forward to show that

$$^*H_{0i} = \sqrt{-g}\,[ijk]H^{jk}, \tag{A9}$$

which is defined as the magnetic field component $H_i$ multiplied by the factor $\sqrt{-g}$.

Equation (A7) can be also written as [16]

$$^*H_{\mu\nu} = \sqrt{-g}\,g_{\mu\lambda}g_{\nu\rho}\,^*F^{\lambda\rho} \tag{A10}$$

where

$$^*F^{\mu\nu} = \frac{1}{2}\varepsilon^{\mu\nu\lambda\rho}F_{\lambda\rho} \tag{A11}$$

with $\quad \varepsilon^{\mu\nu\lambda\rho} = -\dfrac{1}{\sqrt{-g}}[\mu\nu\lambda\rho].$ \hfill (A12)

From, equations (A7) to (A12), we obtain

$$\begin{aligned}\sqrt{-g}\,H_i &= \sqrt{-g}\,\varepsilon_0 g_{00}g_{i\rho}\,^*F^{0\rho} + \sqrt{-g}\,\varepsilon_0 g_{0j}g_{i\rho}\,^*F^{j\rho} \\ &= [jkl]g_{0k}g_{il}\varepsilon_0 E_j + \left(g_{i0}g_{j0} - g_{00}g_{ij}\right)\mu_0^{-1}B_j.\end{aligned} \tag{A13}$$

Or,

$$\begin{aligned}H^i &= \frac{1}{\sqrt{-g}}[jkl]g_{ok}g_{il}\varepsilon_0 E_j + \frac{1}{\sqrt{-g}}\left(g_{io}g_{jo} - g_{00}g_{ij}\right)\mu_0^{-1}B_j \\ &= \beta_{ij}\varepsilon_0 E_j + (\mu^{-1})_{ij}\mu_0^- B_j,\end{aligned} \tag{A14}$$

where

$$\begin{aligned}\beta_{IJ} &= \frac{1}{\sqrt{-g}}[jkl]g_{0k}g_{il}, \\ (\mu^{-1})_{ij} &= \frac{1}{\sqrt{-g}}\left(g_{io}g_{jo} - g_{00}g_{ij}\right)\end{aligned} \tag{A15}$$

The covariant and contravariant components of the metric tensor are related by the following relation:

$$g_{\mu\lambda} g^{\lambda\nu} = \delta_\mu^\nu. \tag{A16}$$

2222

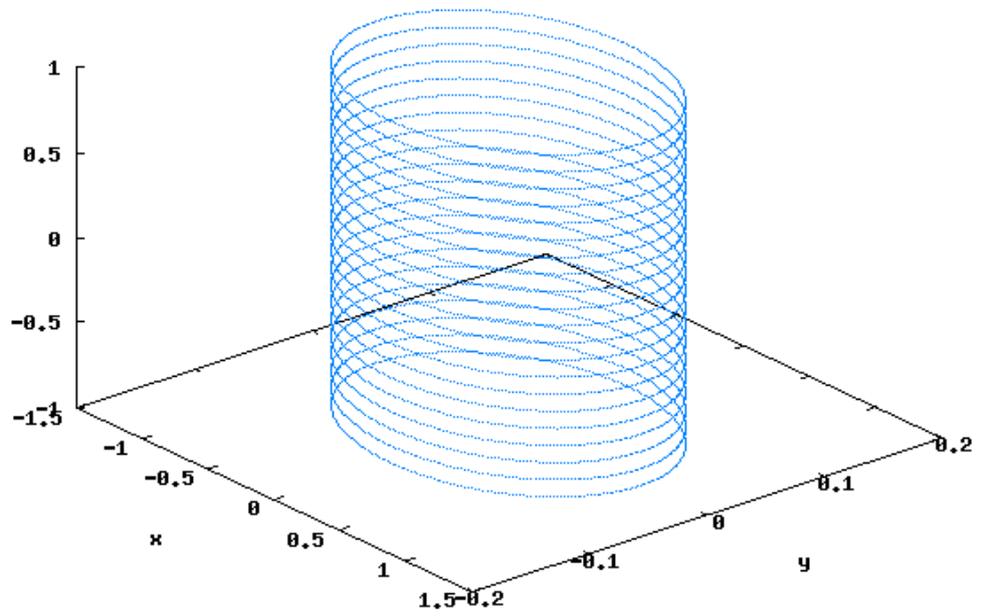

Fig. 1 The boundary of the elliptic cylinder in where the objected is put into for hiding.

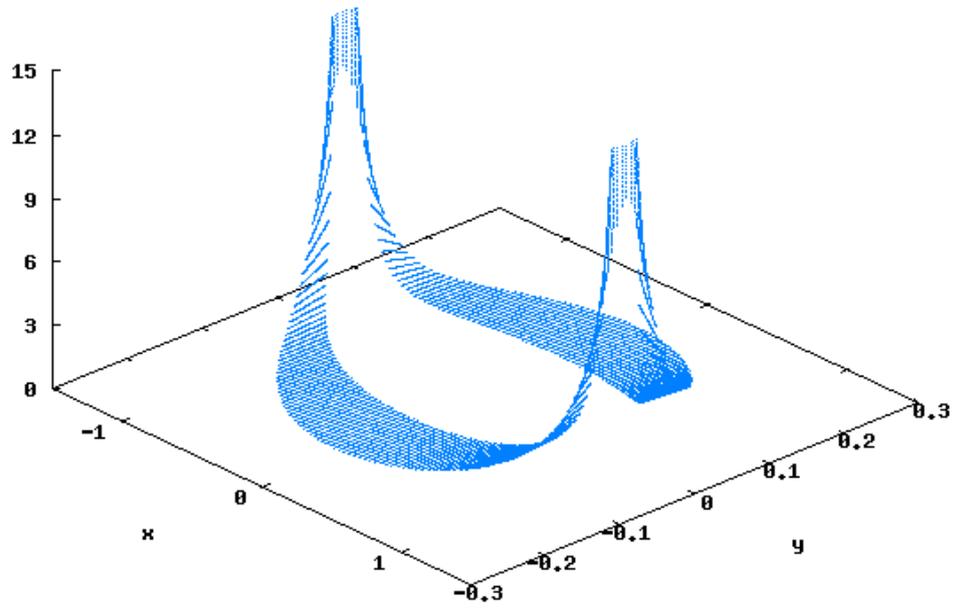

Fig. 2(a) Permittivity tensor $\varepsilon^{u'u'}$ distribution inside the invisibility device of elliptic cylindrical shape.

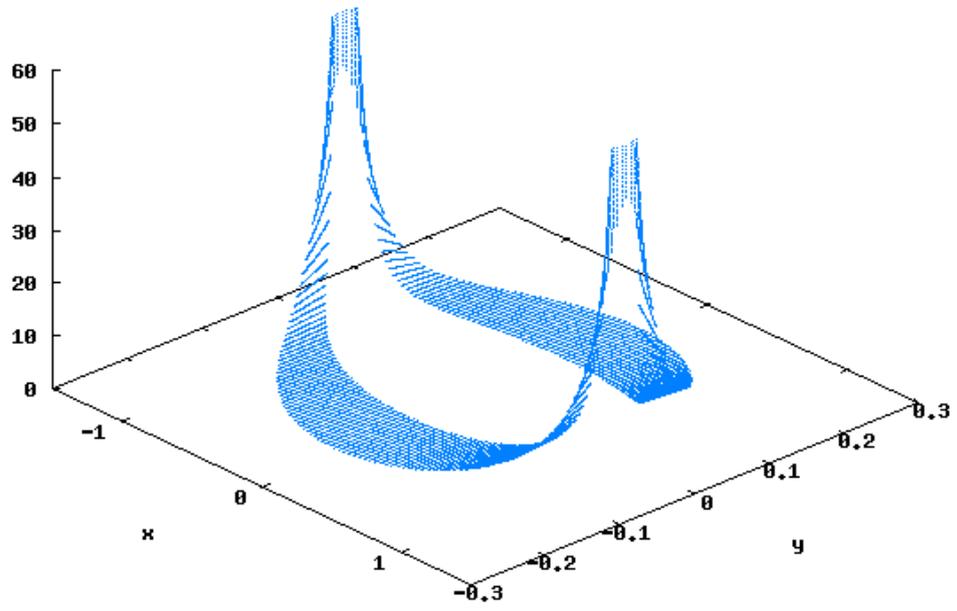

Fig. 2(b) Permittivity tensor $\varepsilon^{vv}$ distribution inside the invisibility device of elliptic cylindrical shape.

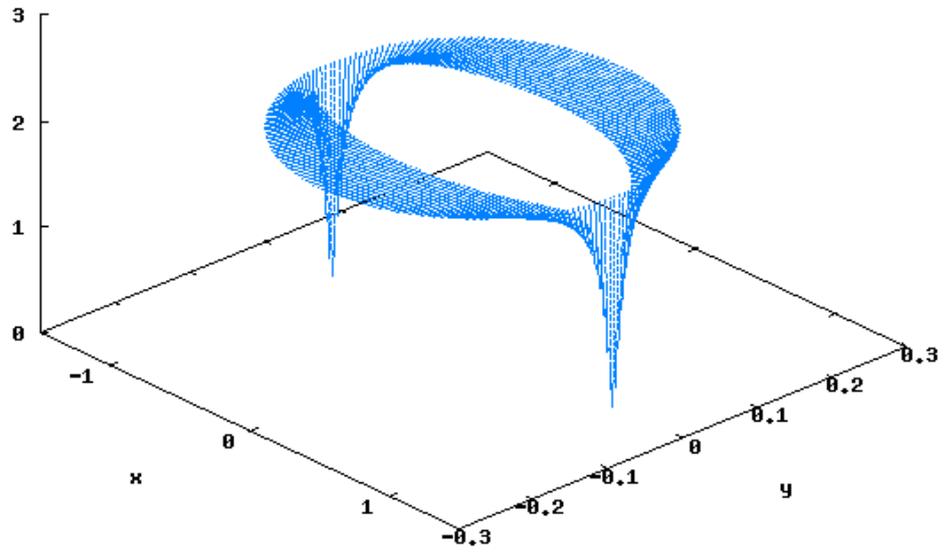

Fig. 2(c) Permittivity tensor $\varepsilon^{zz}$ distribution inside the invisibility device of elliptic cylindrical shape.

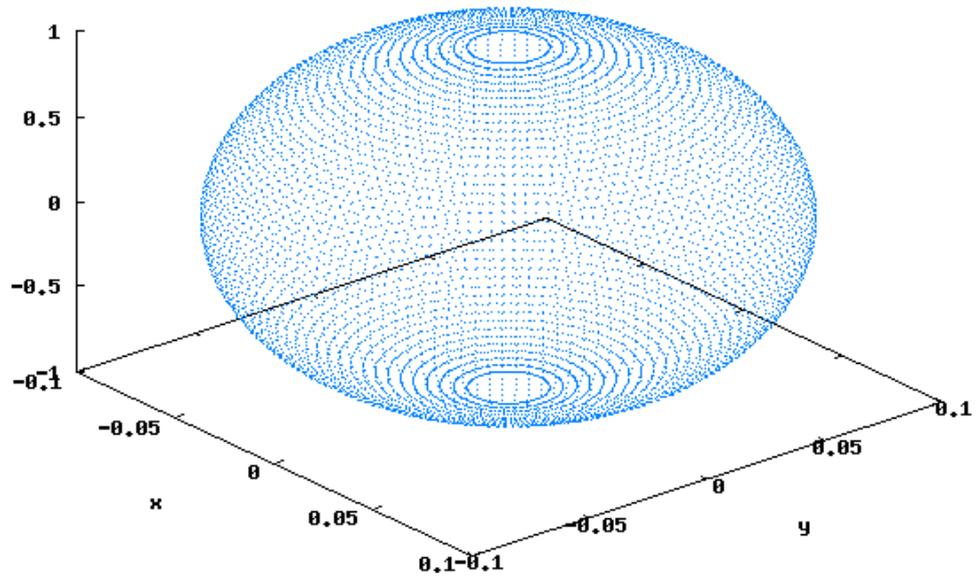

Fig. 3 The boundary of the prolate spheroid in which the objected is put into for hiding.

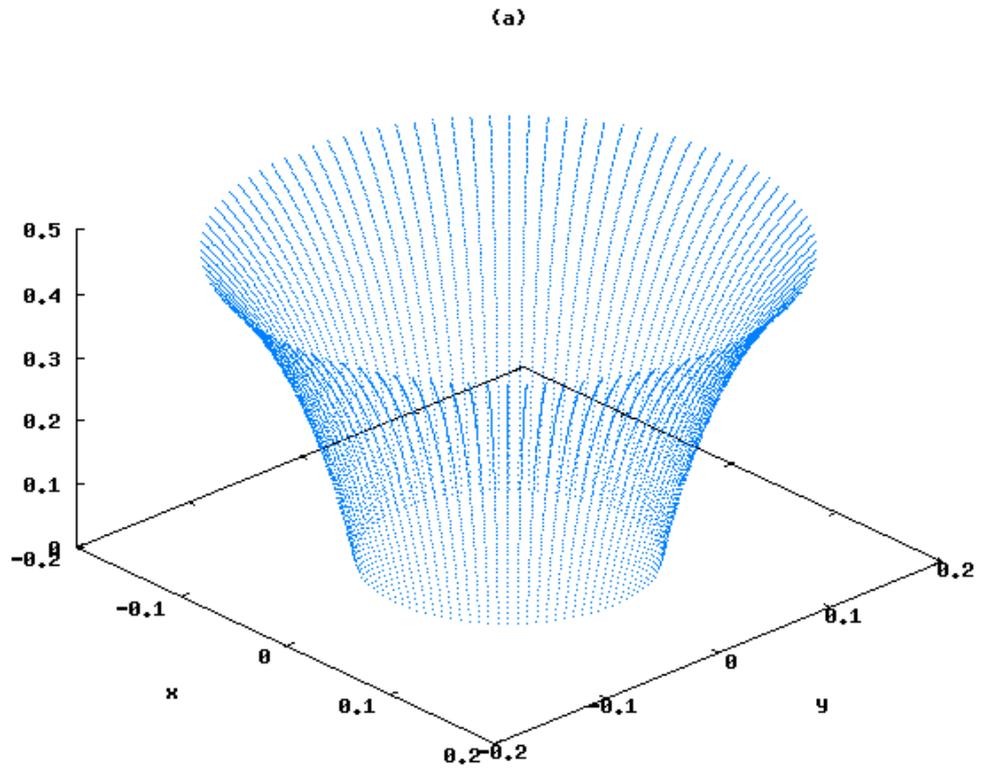

Fig. 4(a) Permittivity tensor $\varepsilon^{uu}$ distribution inside the invisibility device of prolate spheroidal shape.

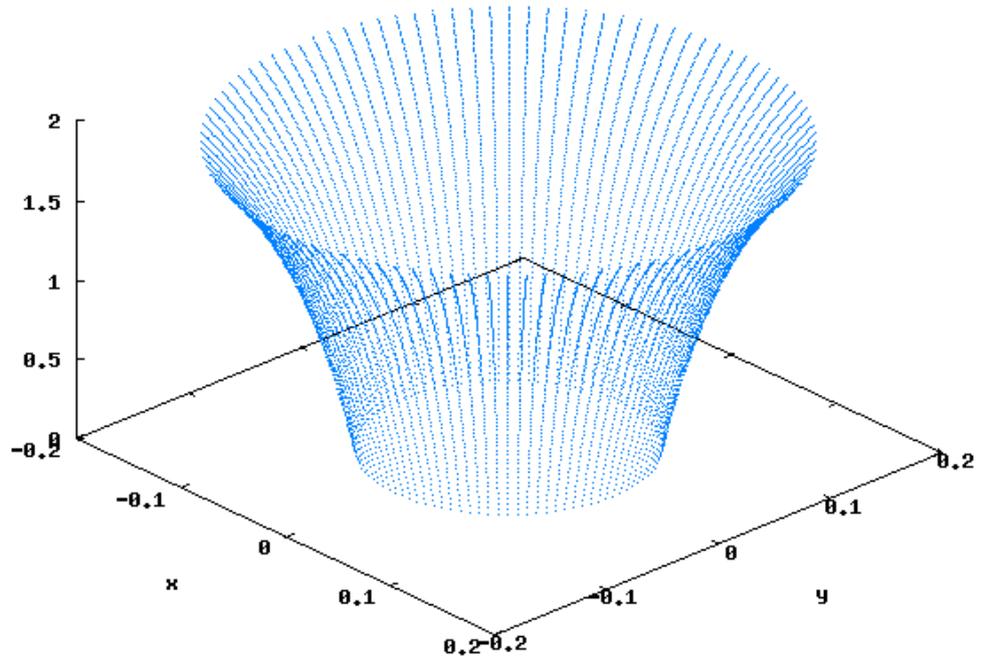

Fig. 4(b) Permittivity tensor $\varepsilon^{vv}$ distribution inside the invisibility device of prolate spheroidal shape.

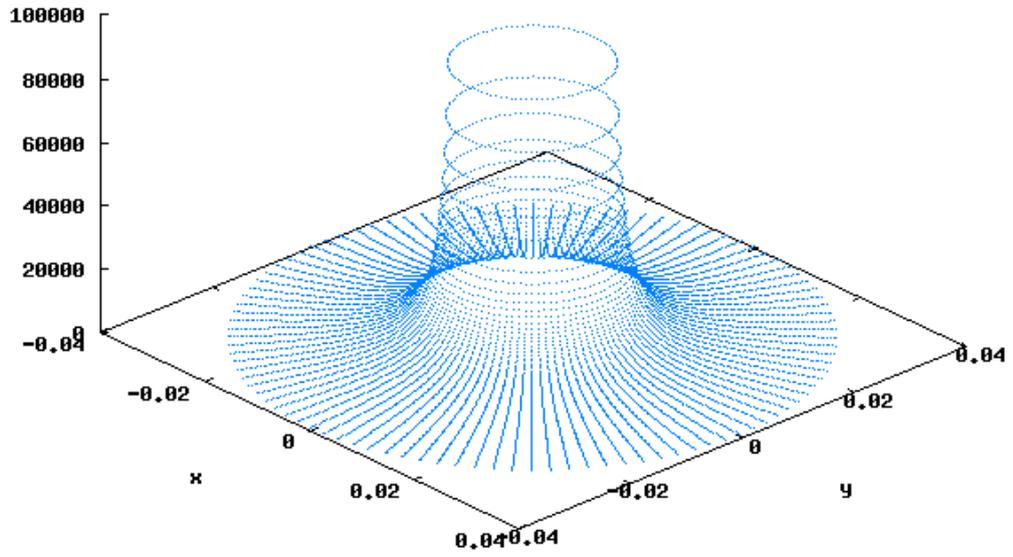

Fig. 4(c) Permittivity tensor $\varepsilon^{\varphi\varphi}$ distribution inside the invisibility device of prolate spheroidal shape.

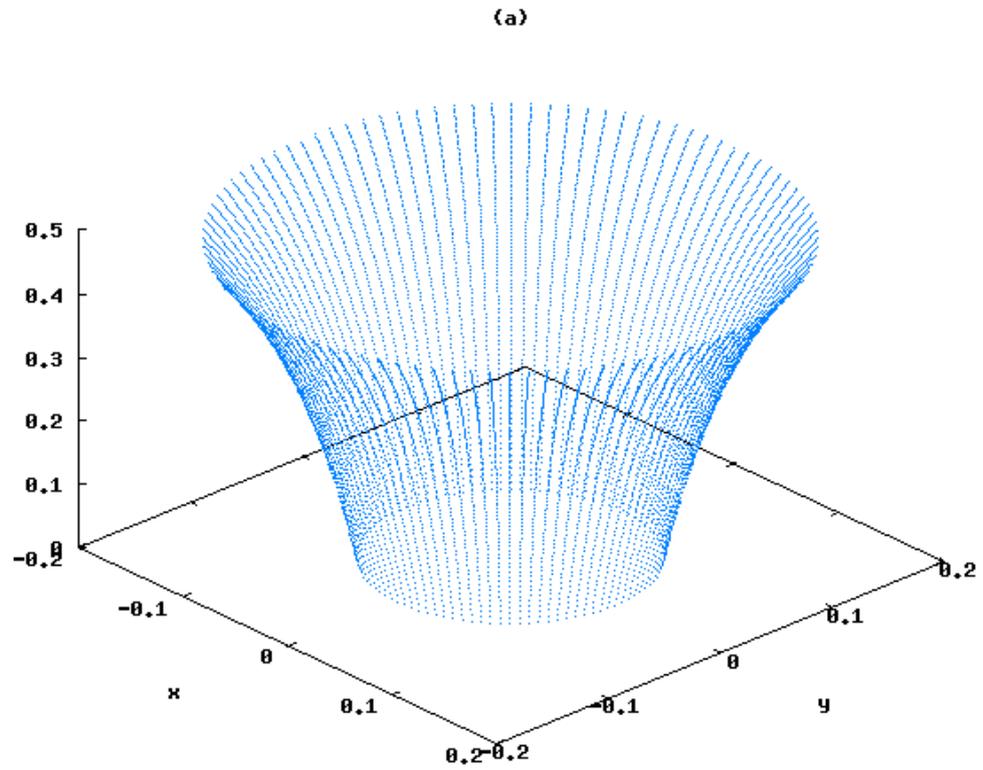

Fig. 5(a) Mixed permittivity tensor $\varepsilon_u^u$ distribution inside the invisibility device of prolate spheroidal shape.

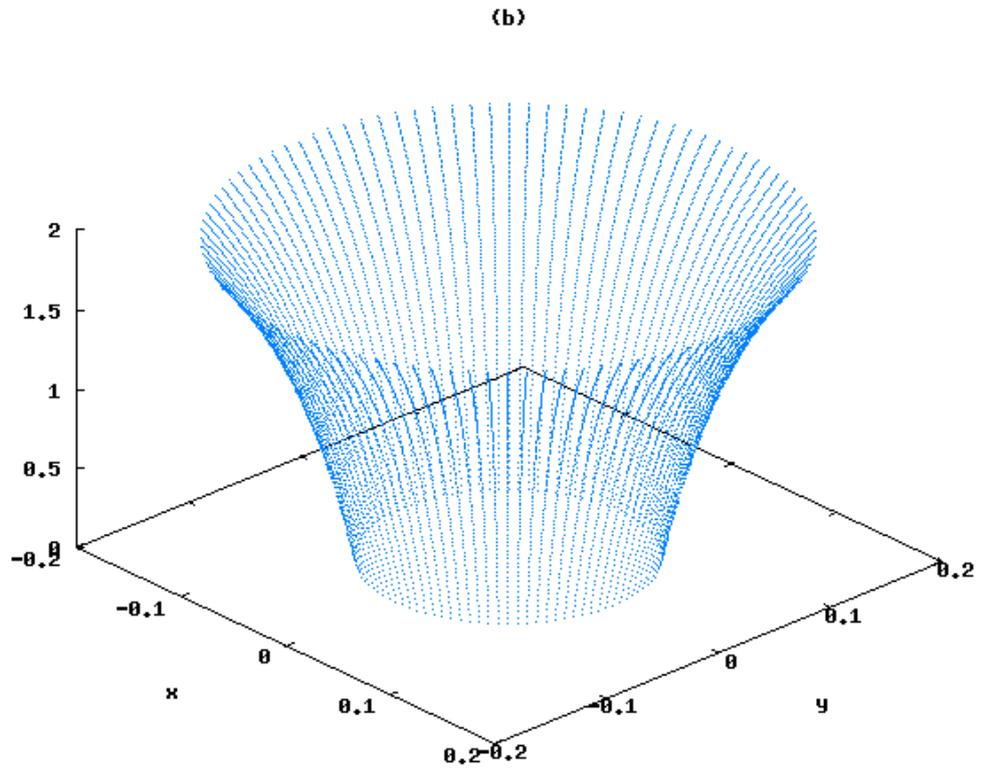

Fig. 5(b) Mixed permittivity tensor $\varepsilon_v^v$ distribution inside the invisibility device of prolate spheroidal shape.

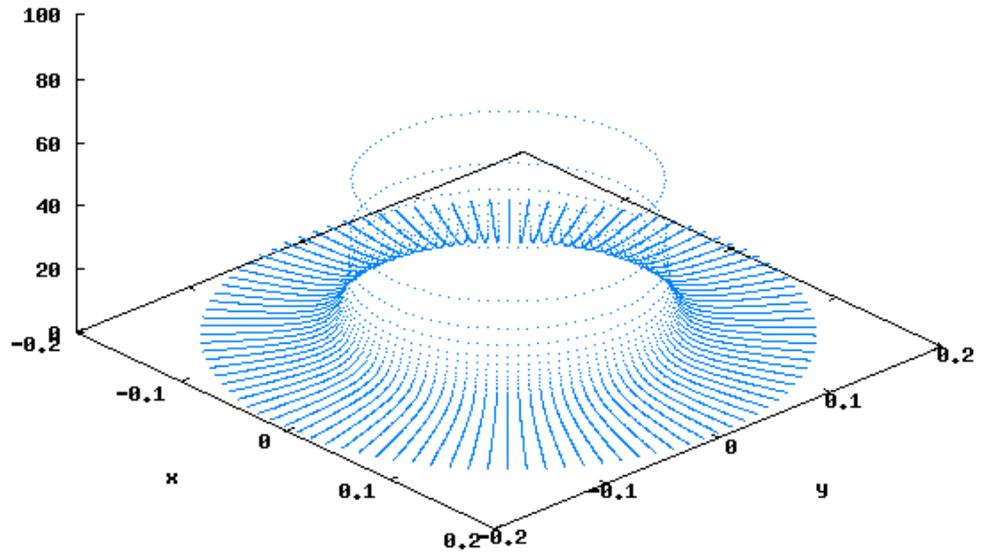

Fig. 5(c) Mixed permittivity tensor $\varepsilon^{\varphi}_{\varphi}$ distribution inside the invisibility device of prolate spheroidal shape.

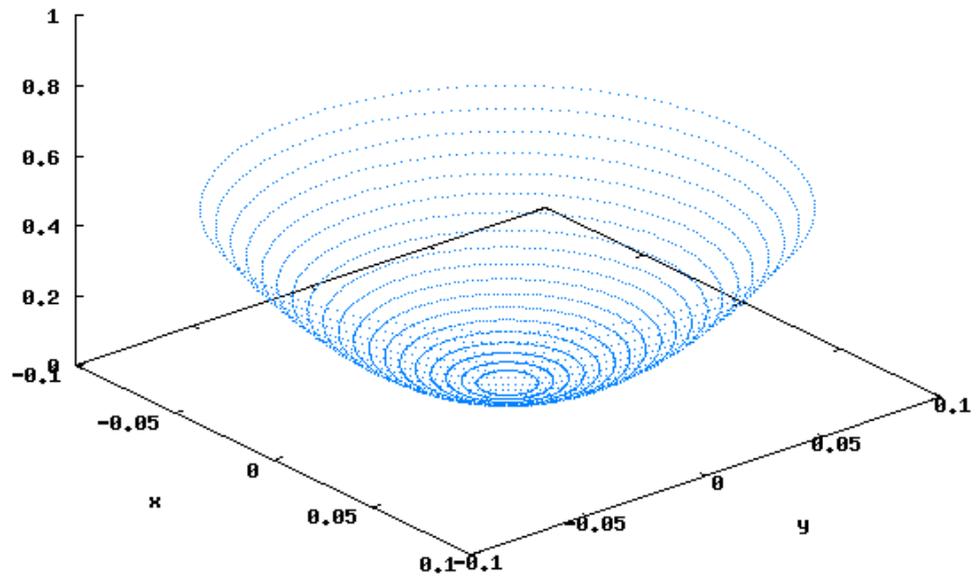

Fig. 6 The boundary of the confocal paraboloid in which the objected is put into for hiding.

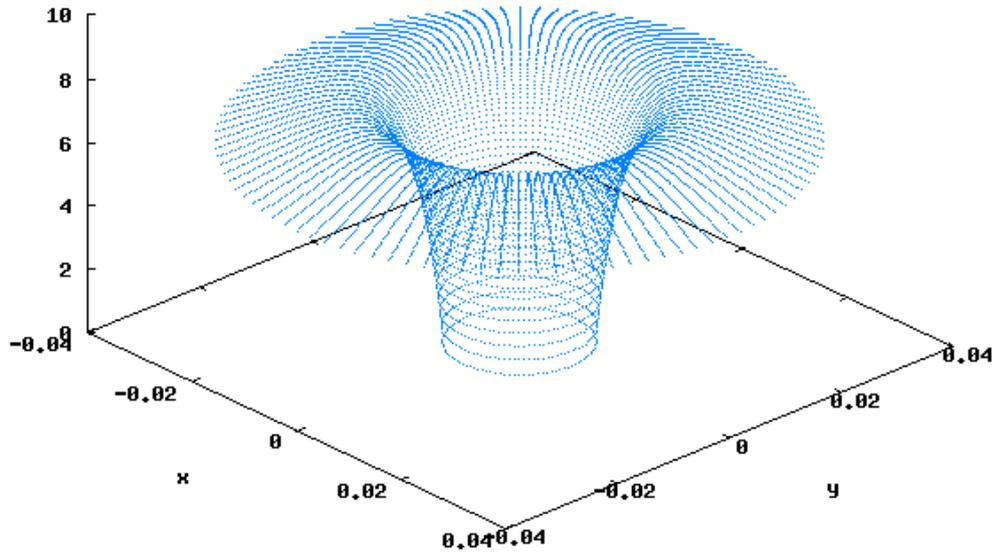

Fig. 7(a) Permittivity tensor $\varepsilon^{\xi\xi}$ distribution inside the invisibility device of confocal paraboloidal shape.

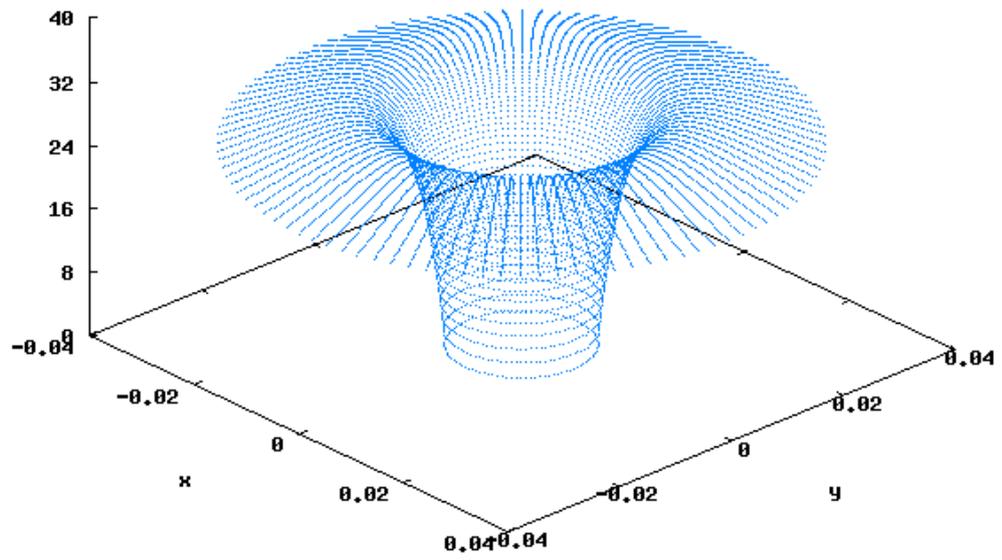

Fig. 7(b) Permittivity tensor $\varepsilon^{\eta\eta}$ distribution inside the invisibility device of confocal paraboloidal shape.

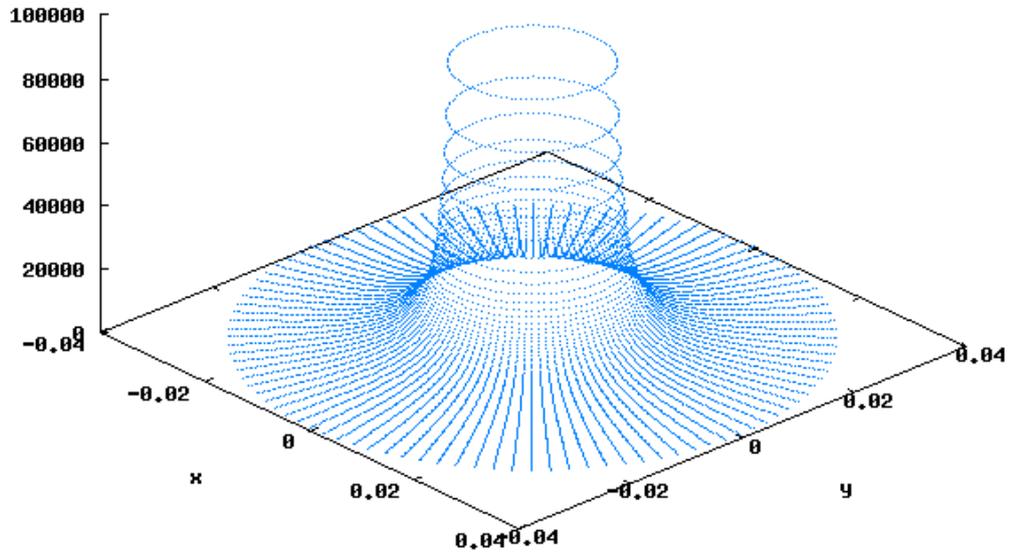

Fig. 7(c) Permittivity tensor $\varepsilon^{\varphi\varphi}$ distribution inside the invisibility device of confocal paraboloidal shape.

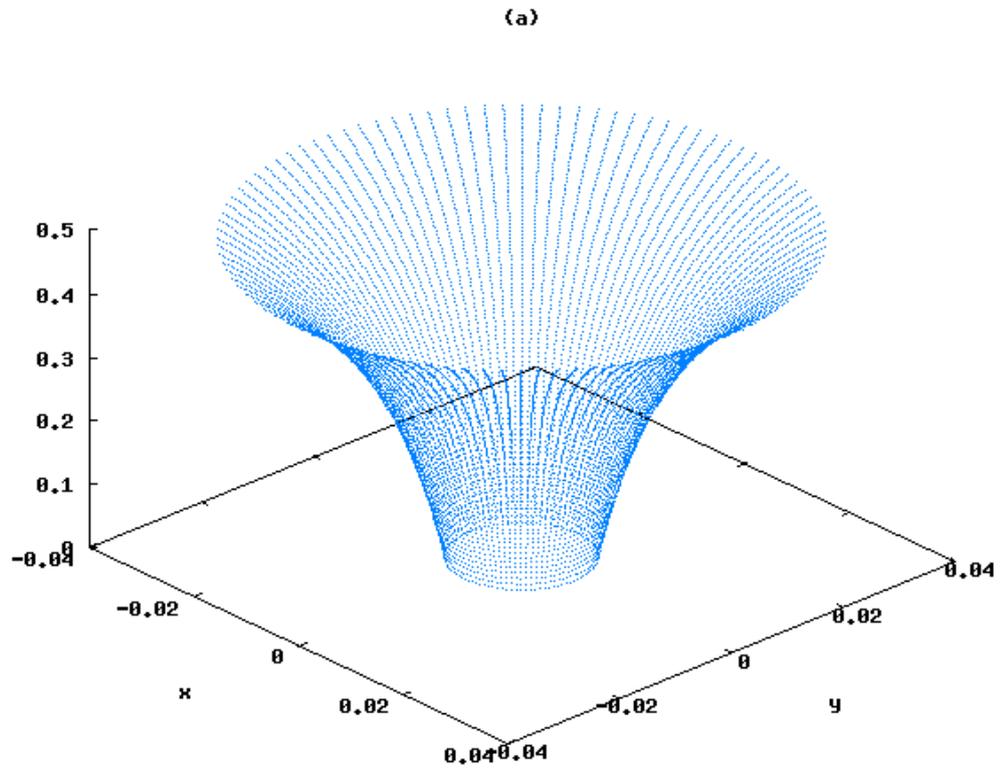

Fig. 8(a) Mixed permittivity tensor $\varepsilon^{\xi}_{\xi}$ distribution inside the invisibility device of confocal paraboloidal shape.

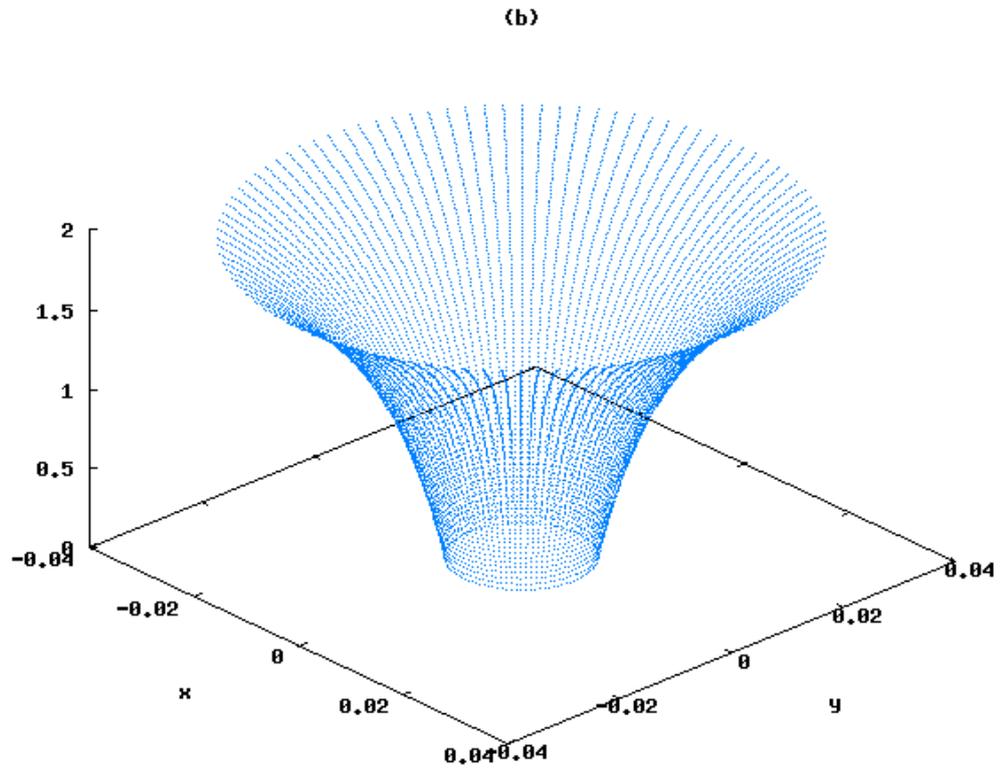

Fig. 8(b) Mixed permittivity tensor $\varepsilon_\eta^\eta$ distribution inside the invisibility device of confocal paraboloidal shape.

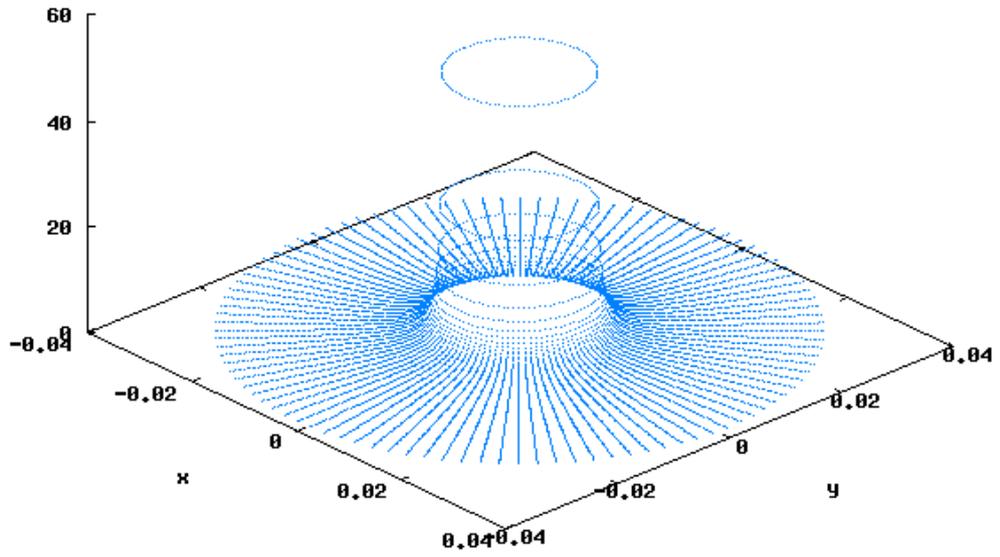

Fig. 8(c) Mixed permittivity tensor $\varepsilon_\varphi^\varphi$ distribution inside the invisibility device of confocal paraboloidal shape.